\documentclass{article}
\usepackage{graphics,bm,amsmath}

\title{On a Supposed Conceptual Inadequacy of the Shannon Information in Quantum Mechanics}
\author{C.G. Timpson\thanks{\tt christopher.timpson@queens.ox.ac.uk}\\The Queen's College, Oxford, OX1 4AW, UK}
\date{18 December 2002}

\begin{document}
\maketitle

\begin{abstract}
Recently, Brukner and Zeilinger (2001) have claimed that the Shannon information is not well defined as a measure of information in quantum mechanics, adducing arguments that seek to show that it is inextricably tied to classical notions of measurement. It is shown here that these arguments do not succeed: the Shannon information does not have problematic ties to classical concepts. In a further argument, Brukner and Zeilinger compare the Shannon information unfavourably to their preferred information measure, $I(\vec{p})$, with regard to the definition of a notion of `total information content'. This argument is found unconvincing and the relationship between individual measures of information and notions of `total information content' investigated. We close by considering the prospects of Zeilinger's Foundational Principle as a foundational principle for quantum mechanics 
\end{abstract}

\section{Introduction}

What role the concept of information might have to play in the foundations of quantum mechanics is a question that has recently excited renewed interest (Fuchs 2000, 2002; Mermin 2001; Wheeler 1990).
Zeilinger, for example, has put forward an information-theoretic principle which he suggests might serve as a foundational principle for quantum mechanics (Zeilinger 1999),(see Appendix).
As a part of this project, Brukner and Zeilinger (2001) have 
criticised Shannon's (1948) measure of information, the quantity fundamental to the discussion of information in both classical and quantum information theory. They claim that the Shannon information is not appropriate as a measure of information in the quantum context and have proposed in its stead their own preferred quantity and a notion of `total information content' associated with it, which latter is supposed to supplant the von Neumann entropy (Brukner and Zeilinger 1999, 2000a, 2000b).

The main aim in Brukner and Zeilinger (2001) is to establish that the Shannon information is intimately tied to classical notions, in particular, to the preconceptions of classical measurement, and that in consequence it cannot serve as a measure of information in the quantum context. They seek to establish this in two ways. First, by arguing that the Shannon measure only makes sense when we can take there to be a pre-existing sequence of bit values in a message we are decoding, which is not the case in general for measurements on quantum systems (consider measurements on qubits in a basis different from their eigenbasis); and second, by suggesting that Shannon's famous third postulate, the postulate that secures the uniqueness of the form of the Shannon information measure (Shannon, 1948) and has been seen by many as a necessary axiom for a measure of information, is motivated by classical preconceptions and does not apply in general in quantum mechanics where we must consider non-commuting observables.

These two arguments do not succeed in showing that the Shannon information is `intimately tied to the notion of systems carrying properties prior to and independent of observation'(Brukner and Zeilinger 2000b:1), however.
The first is based on too narrow a conception of the meaning of the Shannon information and the second, primarily, on a misreading of what is known as the `grouping axiom'. We shall see that the Shannon information is perfectly well defined and appropriate as a measure of information in the quantum context as well as in the classical. We will begin by reviewing some of the different ways in which the Shannon information may be understood (Section~\ref{interpretation}), before examining this pair of arguments and seeing where they go wrong (Section~\ref{arguments}).

Brukner and Zeilinger have a further argument against the Shannon information (Section~\ref{finalargument}). They suggest it is inadequate because it cannot be used to define an acceptable notion of `total information content'. Equally, they insist, the von Neumann entropy cannot be a measure of information content for a quantum system because it has no general relation to information gain from the measurements that we might perform on a system, save in the case of measurement in the basis in which the density matrix is diagonal. By contrast, for a particular set of measurements, their preferred information measure sums to a unitarily invariant quantity that they interpret as `information content', this being one of their primary reasons for adopting this specific measure. This property will be seen to have a simple geometric explanation in the Hilbert-Schmidt representation of density operators however, rather than being of any great information theoretic significance; and this final argument found unpersuasive, as the proposed constraint on any information measure regarding the definition of `total information content' seems unreasonable. Part of the problem is that information content, total or otherwise, is not a univocal concept and we need to be careful to specify precisely what we might mean by it in any given context.

\section{Interpretation of the Shannon information\label{interpretation}} 
The technical concept of information relevant to our discussion, the Shannon information, finds its home in the context of communication theory. We are concerned with a notion of \textit{quantity} of information; and the notion of quantity of information is cashed out in terms of the resources required to transmit messages (which is, note, a very limited sense of quantity).
We shall highlight two main ways in which the Shannon information may be understood, the first of which rests explicitly on Shannon's 1948 noiseless coding theorem.

\subsection{The communication channel}
It is instructive to begin with a quotation of Shannon's:
\begin{quote}
The fundamental problem of communication is that of reproducing at one point either exactly or approximately a message selected at another point. Frequently these messages have \textit{meaning}...These semantic aspects of communication are irrelevant to the engineering problem. (Shannon 1948: 31)
\end{quote}
The communication system consists of an information source, a transmitter or encoder, a (possibly noisy) channel, and a receiver (decoder). It must be able to deal with \textit{any} possible message produced (a string of symbols selected in the source, or some varying waveform), hence it is quite irrelevant whether what is actually transmitted has any meaning or not.

It is crucial to realise that `information' in Shannon's theory is not associated with individual messages, \textit{but rather characterises the source of the messages}. The point of characterising the source is to discover what capacity is required in a communications channel to transmit all the messages the source produces; and it is for this that the concept of the Shannon information is introduced. The idea is that the statistical nature of a source can be used to reduce the capacity of channel required to transmit the messages it produces (we shall restrict ourselves to the case of discrete messages for simplicity).

Consider an ensemble $X$ of letters $\{ x_{1},x_{2}, \ldots ,x_{n} \}$ occurring with probabilities $p_{i}$. This ensemble is our source, from which messages of $N$ letters are drawn. We are concerned with messages of very large $N$. For such messages, we know that typical sequences of letters will contain $Np_{i}$ of letter $x_{i}$, $Np_{j}$ of $x_{j}$ and so on. The number of distinct typical sequences of letters is then given by \[ \frac{N!}{Np_{1}!Np_{2}! \ldots Np_{n}!} \] and using Stirling's approximation, this becomes $2^{NH(X)}$ where
\begin{equation}
H(X)=-\sum_{i=1}^{n}\!p_{i}\log p_{i} \label{shannon}
\end{equation}
is the Shannon information (logarithms are to base 2 to fix the units of information as binary bits).

Now as $N \rightarrow \infty$, the probability of an atypical sequence appearing becomes negligible and we are left with only $2^{NH(X)}$ equiprobable typical sequences which need ever be considered as possible messages. We can thus replace each typical sequence with a binary code number of $NH(X)$ bits and send that to the receiver rather than the original message of $N$ letters ($N\log n$ bits).

The message has been compressed from $N$ letters to $NH(X)$ bits ($\leq N\log n$ bits). Shannon's noiseless coding theorem, of which this is a rough sketch, states that this represents the optimal compression (Shannon 1948). The Shannon information is, then, appropriately called a measure of information because it represents the maximum amount that messages consisting of letters drawn from an ensemble $X$ can be compressed.

One may also make the derivative statement that the information \textit{per letter} in a message is $H(X)$ bits, which is equal to the information of the source. But `derivative' is an important qualification: we can only consider a letter $x_{i}$ drawn from an ensemble $X$ to have associated with it the information $H(X)$ if we consider it to be a member of a typical sequence of $N$ letters, where $N$ is large, drawn from the source. 

Note also that we must strenuously resist any temptation to conclude that because the Shannon information tells us the maximum amount a message drawn from an ensemble can be compressed, that it therefore tells us the irreducible meaning content of the message, specified in bits, which somehow possess their own intrinsic meaning. This idea rests on a failure to distinguish between a code, which has no concern with meaning, and a language, which does (cf. Timpson (2000), Chpt.5).

\subsection{Information and Uncertainty}

Another way of thinking about the Shannon information is as a measure of the amount of information that we \textit{expect} to gain on performing a probabilistic experiment.
The Shannon measure is a measure of the uncertainty of a probability distribution as well as serving as a measure of information. A measure of uncertainty is a quantitative measure of the lack of concentration of a probability distribution; this is called an uncertainty because it measures our uncertainty about what the outcome of an experiment completely described by the probability distribution in question will be. Uffink (1990) provides an axiomatic characterisation of measures of uncertainty, deriving a general class of measures of which the Shannon information is one (see also Maassen and Uffink 1989).

Imagine a truly random probabilistic experiment described by a probability distribution $\vec{p}=\{p_{1},\ldots,p_{n}\}$. The intuitive link between uncertainty and information is that the greater the uncertainty of this distribution, the more we stand to gain from learning the outcome of the experiment.
In the case of the Shannon information, this notion of how much we gain can be made more precise.

Some care is required when we ask `how much do we know about the outcome?' for a probabilistic experiment. In a certain sense, the shape of the probability distribution might provide no information about what an individual outcome will actually be, as any of the outcomes assigned non-zero probability can occur. 
However, we can use the probability distribution to put a \textit{value} on any given outcome. If it is a likely one, then it will be no surprise if it occurs so of little value; if an unlikely one, it is a surprise, hence of higher value. A nice measure for the value of the occurrence of outcome $i$ is $-\log p_{i}$, a decreasing function of the probability of the outcome. We may call this the `surprise' information associated with outcome $i$; it measures the value of what we learn from the experiment given that we know the probability distribution for the outcomes.

If the information that we would gain if outcome $i$ were to occur is $-\log p_{i}$, then before the experiment, the amount of information we expect to gain is given by the expectation value of the `surprise' information, $\sum_{i}p_{i}(-\log p_{i})$; and this, of course, is just the Shannon information $H$ of the probability distribution $\vec{p}$. Hence the Shannon information tells us our expected information gain.  

More generally, a crude sketch of how the relationship between uncertainty and expected information gain might be cashed out for the whole class of measures of uncertainty may be given as follows.
What we know given the probability distribution for an experiment is that if the experiment is repeated very many times then the sequence of outcomes we attain will be one of the typical sequences. How much we learn from actually performing the experiments and acquiring one of those sequences, then, will depend on the number of typical sequences; the more there are, the more we stand to gain. Thus for a quantitative measure of how much information we gain from the sequence of experiments we could just count the number of typical sequences (which would give us $NH$, the Shannon information of the sequence), or we could choose any suitably behaved function (e.g. continuous, invariant under relabelling of the outcome probabilities) that increases as the number of typical sequences increases. This property will follow from Schur concavity, which is the key requirement on Uffink's general class of uncertainty measures $U_{r}(\vec{p})$ (for details of the property of Schur concavity, see Uffink (1990), Nielsen (2001) and Section~\ref{differentnotions} below). $NU_{r}(\vec{p})$ then, can be understood as a measure of the amount of information we gain from a long series $N$ of experiments; we get the average or expected information per measurement by dividing by $N$, but note that this quantity only makes sense if we consider the individual measurement as part of a long sequence of measurements.

A precisely similar story can be told for a measure of `how much we know' given a probability distribution. This will be the inverse of an uncertainty: we want a measure of the concentration of a probability distribution; the more concentrated, the more we know about what the outcome will be (which just means, the better we can predict the outcome). A function that decreases as the number of typical sequences increases (Schur convexity) will give our quantitative measure of how much we know about what the outcome of a long run of experiments will be: the more typical sequences there are the less we know. We are again talking of a long run of experiments, so `how much we know' for a single experiment will only make sense as an average value, when the single experiment is considered as a member of a long sequence. So note again that to say we have a certain amount of information (knowledge) about what the outcome of an experiment will be is not to claim that we have partial knowledge of some predetermined fact about the outcome of an experiment.

\subsection{The minimum number of questions needed to specify a sequence}

The final common interpretation of the Shannon information is as the minimum average number of binary questions needed to specify a sequence drawn from an ensemble (Uffink 1990; Ash 1965), although this appears not to provide an interpretation of the Shannon information actually independent of the previous two.

Imagine that a long sequence $N$ of letters is drawn from the ensemble $X$, or that $N$ independent experiments whose possible outcomes have probabilities $p_{i}$ are performed, but the list of outcomes is kept from us. Our task is to determine what the sequence is by asking questions to which the guardian of the sequence can only answer `yes' or `no'; and we choose to do so in such a manner as to minimize the average number of questions needed. We need to be concerned with the \textit{average} number to rule out lucky guesses identifying the sequence.

If we are trying to minimize the average number of questions, it is evident that the best questioning strategy will be one that attempts to rule out half the possibilities with each question, for then whatever the answer turns out to be, we still get the maximum value from each question. Given the probability distribution, we may attempt to implement this strategy by dividing the possible outcomes of each individual experiment into classes of equal probability, and then asking whether or not the outcome lies in one of these classes. We then try and repeat this process, dividing the remaining set of possible outcomes into two sets of equal probabilities, and so on. It is in general not possible to proceed in this manner, dividing a finite set of possible outcomes into two sets of equal probabilities, and it can be shown that in consequence the average number of questions required if we ask about each individual experiment in isolation is greater than or equal to $H(X)$. However, if we consider the $N$ repeated experiments, where $N$ tends to infinity, and consider asking joint questions about what the outcomes of the independent experiments were, we can always divide the classes of possibilities of (joint) outcomes in the required way. Now we already know that for large $N$, there are $2^{NH(X)}$ typical sequences, so given that we can strike out half the possible sequences with each question, the minimum average number of questions needed to identify the sequence is $NH(X)$. (These last results are again essentially the noiseless coding theorem.)

It is not immediately obvious, however, why the minimum average number of questions needed to specify a sequence should be related to the notion of information. (Again, the tendency to think of bits and binary questions as irreducible meaning elements is to be resisted.) It seems, in fact that this is either just another way of talking about the maximum amount that messages drawn from a given ensemble can be compressed, in which case we are back to the interpretation of the Shannon information in terms of the noiseless coding theorem, or it is providing a particular way of characterising how much we stand to gain from learning a typical sequence, and we return to an interpretation in terms of our expected information gain.

\section{Two arguments against the Shannon information\label{arguments}}
With this preamble behind us, we may turn to the first of the arguments against the Shannon information.
\subsection{Are pre-existing bit-values required?}\label{preexistingbits}
Since the quantity $-\sum_{i}{p_{i}}\log p_{i}$ is meaningful for any (discrete) probability distribution (and can be generalised for continuous distributions), Brukner and Zeilinger's argument must be that when we have probabilities arising from measurements on quantum systems, $-\sum_{i}{p_{i}}\log p_{i}$ does not correspond to a concept of \textit{information}. Their argument concerns measurements on systems that are all prepared in a given state $|\psi\rangle$, where $|\psi\rangle$ may not be an eigenstate of the observable we are measuring. The probability distribution $\vec{p}=\{p_{1},\ldots,p_{n}\}$ for measurement outcomes will be given by $p_{i}={\mbox Tr}(|\psi\rangle\langle \psi |  P_{i})$, where $P_{i}$ are the operators corresponding to different measurement outcomes (projection operators in the spectral decomposition of the observable, for projective measurements). 

Brukner and Zeilinger suggest that the Shannon information has no meaning in the quantum case, because the concept lacks an `operational definition' in terms of the number of binary questions needed to specify an actual concrete sequence of outcomes. In general in a sequence of measurements on quantum systems, we cannot consider there to be a pre-existing sequence of possessed values, at least if we accept the orthodox eigenvalue-eigenstate link for the ascription of definite values (see e.g. Bub (1997))\footnote{In a footnote, Brukner and Zeilinger suggest that the Kochen-Specker theorem in particular raises problems for the operational definition of the Shannon information. It is not clear, however, why the impossibility of assigning \textit{context independent} yes/no answers to questions asked of the system should be a problem if we are considering an \textit{operational} definition. Presumably such a definition would include a concrete specification of the experimental situation, i.e. refer to the context, and then we are not concerned with assigning a value to an \textit{operator} but to the outcome of a specified experimental procedure, and this can be done perfectly consistently, if we so wish. The de-Broglie Bohm theory, of course, provides a concrete example (Bell 1982).\label{KS}}, and this rules out, they insist, interpreting the Shannon measure as an amount of information:
\begin{quote}
The nonexistence of well-defined bit values prior to and independent of observation suggests that the Shannon measure, as defined by the number of binary questions needed to determine the particular \textit{observed} sequence 0's and 1's, becomes problematic and even untenable in defining our uncertainty as given \textit{before} the measurements are performed. (Brukner and Zeilinger 2001:1)\vspace{\baselineskip}\\
...No definite outcomes exist before measurements are performed and therefore the number of different possible sequences of outcomes does not characterize our uncertainty about the individual system before measurements are performed. (Brukner and Zeilinger 2001:3) 
\end{quote}
These two statements should immediately worry us, however.
Recall the key points of the interpretation of the Shannon information: given a long message (a long run of experiments), we \textit{know} that it will be one of the typical sequences that is instantiated. Given $\vec{p}$, we can say what the typical sequences will be, how many there are, and hence the number of bits ($NH(X)$) needed to specify them, 
independent of whether or not there is a pre-existing sequence of bit values.
It is irrelevant whether there already is some concrete sequence of bits or not; all possible sequences that will be produced will require the same number of bits to specify them as any sequence produced will always be one of the typical sequences. It clearly makes no difference to this whether the probability distribution is given classically or comes from the trace rule.
Also, the number of different possible sequences does indeed tell us about our uncertainty before measurement: what we know is that one of the typical sequences will be instantiated, what we are ignorant of is which one it will be, and we can put a measure on how ignorant we are simply by counting the number of different possibilities. Brukner and Zeilinger's attempted distinction between uncertainty before and after measurement is not to the point, the uncertainty is a function of the probability distribution and this is perfectly well defined before measurement\footnote{We may need to enter at this point the important note that the Shannon information is not supposed to describe our \textit{general} uncertainty when we know the \textit{state}, this is a job for a measure of mixedness such as the von Neumann entropy, see below.}. 

Brukner and Zeilinger have assumed that it is a necessary and sufficient condition to understand $H$ as a measure of information that there exists some concrete string of $N$ values, for then and only then can we talk of the minimum number of binary questions needed to specify the string.
But as we have now seen, it is \textit{not} a necessary condition that there exist such a sequence of outcomes.

We are not in any case forced to assume that $H$ is about the number of questions needed to specify a sequence in order to understand it as a measure of information; we also have the interpretations in terms of the maximum amount a message drawn from an ensemble described by the probability distribution $\vec{p}$ can be compressed, and as the expected information gain on measurement. (And as we have seen, one of these two interpretations must in fact be prior.)
Furthermore, the absence of a pre-existing string need not even be a problem for the minimum average questions interpretation --- we can ask about the minimum average number of questions that \textit{would} be required if we \textit{were} to have a sequence drawn from the ensemble. So again, the pre-existence of a definite string of values is not a necessary condition.

It is not a sufficient condition either, because, faced with a string of $N$ definite outcomes, in order to interpret $NH$ as the minimum \textit{average} number of questions needed to specify the sequence, we need to know that we in fact have a typical sequence, that is, we need to imagine an ensemble of such typical sequences and furthermore, to assume that the relative frequencies of each of the outcomes in our actual string is representative of the probabilities of each of the outcomes in the notional ensemble from which the sequence is drawn. If we do not make this assumption, then the minimum number of questions needed to specify the state of the sequence must be $N$ --- we cannot imagine that the statistical nature of the source from which the sequence is notionally drawn allows us to compress the message. So even in the classical case, the concrete sequence on its own is not enough and we need to consider an ensemble, either of typical sequences or an ensemble from which the concrete sequence is drawn. In this respect the quantum and classical cases are completely on a par. The same assumption needs to be made in both cases, namely, that the probability distribution $\vec{p}$, either known in advance, or derived from observed relative frequencies, correctly describes the probabilities of the different possible outcomes.
The fact that no determinate sequence of outcomes exists before measurement does not pose any problems for the Shannon information in the quantum context.

Reiterating their requirements for a satisfactory notion of information, Brukner and Zeilinger say:
\begin{quote}
We require that the information gain be directly based on the observed probabilities, (and not, for example, on the precise sequence of individual outcomes observed on which Shannon's measure of information is based). (Brukner and Zeilinger 2000b:1)
\end{quote}
But as we have seen, it is false that the Shannon measure must be based on a precise sequence of outcomes (this is not a necessary condition) and the Shannon measure already \textit{is} and \textit{must be} based on the observed probabilities (a sequence of individual outcomes on its own is not sufficient).

There is, however, a difference between the quantum and classical cases that Brukner and Zeilinger may be attempting to capture. Suppose we have a sequence of  $N$ qubits that has actually been used to encode some information, that is, the sequence of qubits is a channel to which we have connected a classical information source. For simplicity, imagine that we have coded in orthogonal states. Then the state of the sequence of qubits will be a product of $|{0}\rangle$'s and $|1\rangle$'s and for measurements in the encoding basis, the sequence will have a Shannon information equal to $NH(A)$ where $H(A)$ is the information of the classical source. If we do not measure in the encoding basis, however, the sequence of 0's and 1's we get as our outcomes will differ from the values originally encoded and the Shannon information of the resulting sequence will be greater than that of the original\footnote{We may think of our initial sequence of qubits as forming an ensemble described by the density operator $\rho = p_{1}|0\rangle\langle 0| + p_{2}|1\rangle\langle 1|$, where $p_{1}, p_{2}$ are the probabilities for 0 and 1 in our original classical information source. Any (projective) measurement that does not project onto the eigenbasis of $\rho$ will result in a post-measurement ensemble that is more mixed than $\rho$ (see e.g. Nielsen (2001), Peres (1995) and below) and hence will have a greater uncertainty, thus a greater Shannon information, or any other measure of information gain.}. We have introduced some `noise' by measuring in the wrong basis. The way we describe this sort of situation, though (Schumacher 1995), is to use the Shannon mutual information $H(A:B)=H(A)-H(A|B)$, where $B$ denotes the outcome of measurement of the chosen observable (outcomes $b_{i}$ with probabilities $p(b_{i})$) and the `conditional entropy' $H(A|B)=\sum_{i=1}^{n}p(b_{i})H(p(a_{1}|b_{i}),\ldots,p(a_{m}|b_{i}))$, characterises the noise we have introduced by measuring in the wrong basis. $H(B)$ is the information (per letter) of the sequence that we are left with after measurement, $H(A:B)$ tells us the amount of information that we have actually managed to transmit down our channel, i.e. the amount (per letter) that can be decoded when we measure in the wrong basis.

\subsection{The grouping axiom}
The first argument has not revealed any difficulties for the Shannon information in the quantum context, so let us now turn to the second. 

In his original paper, Shannon put forward three properties as reasonable requirements on a measure of uncertainty and showed that the only function satisfying these requirements has the form $H=-K\sum_{i}p_{i}\log p_{i}$.\footnote{In contrast to some later writers, however, notably Jaynes (1957), he set little store by this derivation, seeing the justification of his measure as lying rather in its implications (Shannon 1948). Save the noiseless coding theorem, the most significant of the implications that Shannon goes on to draw are, as has been pointed out by Uffink, consequences of the property of Schur concavity and hence shared by the general class of measures of uncertainty derived in Uffink (1990).}

\begin{figure}[btp]
\begin{center}
\includegraphics*[178,499][321,575]{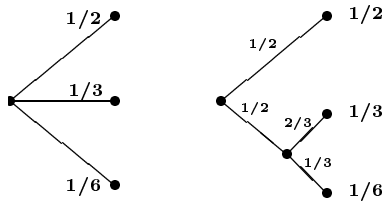}
\end{center}
\caption[Shannon's third requirement]{
The probability distribution $\vec{p}= \{\frac{1}{2},\frac{1}{3},\frac{1}{6}\}$ can be considered as given for the three outcomes directly, or we could consider first a choice of two equiprobable events, followed by a second choice of two events with probabilities $\frac{2}{3},\frac{1}{3}$, conditional on the second, say, of the first two events occurring, a `decomposition' of a single choice into two successive choices, the latter of which will only be made half the time. Shannon's third requirement says that the uncertainty in $\vec{p}$ will be given by $H(\frac{1}{2},\frac{1}{3},\frac{1}{6})= H(\frac{1}{2},\frac{1}{2}) + \frac{1}{2}H(\frac{2}{3},\frac{1}{3})$: the uncertainty of the overall choice is equal to the uncertainty of the first stage of the choice, plus the uncertainty of the second choice weighted by its probability of occurrence.}\label{decomposition}
\end{figure}

The first two requirements are that $H$ should be continuous in the $p_{i}$ and that for equiprobable events ($p_{i}=1/n$), $H$ should be a monotonic increasing function of $n$.
The third requirement is the strongest and the most important in the uniqueness proof. It states that if a choice is broken down into two successive choices, the original $H$ should be a weighted sum of the individual values of $H$. The meaning of this rather non-intuitive constraint is usually demonstrated with an example (see Fig. \ref{decomposition}).
A precise statement of Shannon's third requirement (one that includes also the second requirement as a special case) is due to Faddeev (1957) and is often known as the Faddeev grouping axiom:
\newtheorem{groupingaxiom}{Grouping Axiom}
\begin{groupingaxiom}[Faddeev] \label{faddeevgroupingaxiom}
For every $n\geq2$
\begin{equation}
H(p_{1},p_{2},\ldots,p_{n-1},q_{1},q_{2}) = H(p_{1},\ldots,p_{n-1},p_{n}) + p_{n}H(\frac{q_{1}}{p_{n}},\frac{q_2}{p_{n}})
\end{equation}
where $p_{n}=q_{1}+q_{2}$.
\end{groupingaxiom} 
The form of the Shannon information follows uniquely from requiring $H(p, 1-p)$ to be continuous for $0\leq p \leq 1$ and positive for at least one value of $p$, permutation invariance of $H$ with respect to relabelling of the $p_{i}$, and the grouping axiom.

`Grouping axiom' is an appropriate name. As it is standardly understood (see e.g. Ash (1965), Uffink (1990), Jaynes (1957)), we consider that instead of giving the probabilities $p_{1},\ldots,p_{n}$ of the outcomes $x_{1},\ldots,x_{n}$ of a probabilistic experiment directly, we may imagine grouping the outcomes into composite events (whose probabilities will be given by the sum of the probabilities of their respective component events), and then specifying the probabilities of the outcome events conditional on the occurrence of the composite events to which they belong; this way of specifying the probabilistic experiment being precisely equivalent to the first. So we might group the first $k$ events together into an event $A$, which would have a probability $p(A)=\sum_{i=1}^{k}p_{i}$, and the remaining $n-k$ into an event $B$ of probability $p(B)=\sum_{i=k+1}^{n}p_{i}$; and then give the conditional probabilities of the events $x_{1},\ldots,x_{k}$ conditional on composite event $A$ occurring, $(p_{1}/p(A)),\ldots,(p_{k}/p(A))$, and similarly the conditional probabilities for the events $x_{k+1},\ldots,x_{n}$ conditional on event $B$. The grouping axiom then concerns how the uncertainty measures should be related for these different descriptions of the same probabilistic experiment. It says that our uncertainty about which event will occur should be equal to our uncertainty about which group it will belong to plus the expected value of the uncertainty that would remain if we were to know which group it belonged to (this expected value being the weighted sum of the uncertainties of the conditional distributions, with weights given by the probability of the outcome lying within a given group).

So in particular, let us imagine an experiment with $n+1$ outcomes which we label $a_{1}, a_{2},\ldots, a_{n-1},b_{1},b_{2}$, having probabilities $p_{1},\ldots,p_{n-1},q_{1},q_{2}$ respectively. We can define an event $a_{n}=b_{1}\cup b_{2}, b_{1}\cap b_{2}=\emptyset$, which would have probability $p_{n}=q_{1}+q_{2}$ and the probabilities for $b_{1}$ and $b_{2}$ conditional on $a_{n}$ occurring will then be $\frac{q_{1}}{p_{n}}, \frac{q_{2}}{p_{n}}$ respectively. Grouping Axiom \ref{faddeevgroupingaxiom} says that the uncertainty in the occurrence of events $a_{1}, a_{2},\ldots, a_{n-1},b_{1},b_{2}$ is equal to the uncertainty for the occurrence of events $a_{1},\ldots,a_{n}$ plus the uncertainty for the occurrence of $b_{1},b_{2}$ conditional on $a_{n}$ occurring, weighted by the probability that $a_{n}$ should occur. 

Brukner and Zeilinger suggest that the grouping axiom, however, embodies certain classical presumptions that do not apply in quantum mechanics. This entails that the axiomatic derivation of the form of the Shannon measure does not go through and that the Shannon information ceases to be a measure of uncertainty in the quantum context.
The argument turns on their interpretation of the grouping axiom, which differs from the standard interpretation in that it refers to joint experiments.

\subsubsection{Brukner and Zeilinger's interpretation}
\begin{sloppypar}
If we take an experiment, $A$, with outcomes $a_{1},\ldots,a_{n}$ and probabilities $(p(a_{1}),\ldots,p(a_{n})) = (p_{1},\ldots,p_{n})$ and an experiment, $B$, with outcomes $b_{1},b_{2}$, then for the joint experiment $A \wedge B$, the event $a_{n}$ is the union of the two disjoint events $a_{n}\wedge b_{1}$ and $a_{n}\wedge b_{2}$. Let us assign to these two events the probabilities $q_{1}$ and $q_{2}$ respectively. Then $p(a_{n})=p(a_{n}\wedge b_{1})+p(a_{n}\wedge b_{2})=q_{1}+q_{2}=p_{n}$. On this interpretation, the left hand side of Grouping Axiom \ref{faddeevgroupingaxiom} is to be understood as denoting the uncertainty in the experiment with outcomes $a_{1},a_{2},\ldots,a_{n-1},a_{n}\wedge b_{1},a_{n}\wedge b_{2}$.
\end{sloppypar}

If $a_{n}$ occurs, the conditional probabilities for $b_{1},b_{2}$ will be $p(a_{n}\wedge b_{1})/p(a_{n})=q_{1}/p_{n}, p(a_{n}\wedge b_{2})/p(a_{n})=q_{2}/p_{n}$ respectively, and so $H(\frac{q_{1}}{p_{n}},\frac{q_{2}}{p_{n}})$ is the uncertainty in the value of $B$ given that $a_{n}$ occurs.

The grouping axiom can now be rewritten as:
\begin{groupingaxiom}[Brukner and Zeilinger]\label{BZ1} \mbox{ }
\begin{multline}
H \left(p(a_{1}),p(a_{2}),\ldots,p(a_{n-1}),p(a_{n}\wedge b_{1}),p(a_{n}\wedge b_{2})\right )\\
= H\left(p(a_{1}),p(a_{2}),\ldots,p(a_{n})\right) 
  + p(a_{n})H\left(p(b_{1}|a_{n}),p(b_{2}|a_{n})\right).
\end{multline}
\end{groupingaxiom}
Generalizing to the case in which we have $m$ outcomes for experiment $B$ and distinguish $B$ values for all $n$ $A$ outcomes, so that we have $mn$ outcomes $a_{i}\wedge b_{j}$, the grouping axiom becomes:
\begin{groupingaxiom}[Brukner and Zeilinger]\label{BZ2} \mbox{ }
\[H(A\wedge B)=H(A)+H(B|A)\]
\end{groupingaxiom}
From the point of view of Shannon's original presentation, this expression appears as a theorem rather than an axiom, being a consequence of the logarithmic form of the Shannon information and the definition of the conditional entropy.

\subsubsection{The inapplicability argument}

The classical assumptions made explicit, Brukner and Zeilinger suggest, in Grouping Axioms \ref{BZ1} and \ref{BZ2} are that attributes corresponding to all possible measurements can be assigned to a system simultaneously (in this case, $a_{i}, b_{j}$ and $a_{i}\wedge b_{j}$); and that measurements can be made ideally non-disturbing. Grouping Axiom \ref{BZ2}, for example, is supposed to express the fact that classically, the information we expect to gain from a joint experiment $A\wedge B$, is the same as the information we expect to gain from first performing $A$, then performing $B$ (where the uncertainty in $B$ is updated conditional on the $A$ outcome, but our ability to predict $B$ outcomes is not degraded by the $A$ measurement).

Their inapplicability argument is simply that as the grouping axiom requires us to consider joint experiments, the uniqueness proof for the Shannon information will fail in the quantum context, because we can consider measurements of non-commuting observables and the joint probabilities on the left hand side of Grouping Axiom \ref{BZ1} will not be defined for such observables; thus the grouping axiom will fail to hold. Furthermore, the grouping axiom shows that the Shannon information embodies classical assumptions, so the Shannon measure will not be justified as a measure of uncertainty because these assumptions do not hold in the quantum case. The result is that
\begin{quote}
...only for the special case of commuting, i.e., simultaneously definite observables, is the Shannon measure of information applicable \textit{and the use of the Shannon information justified to define the uncertainty given before quantum measurements are performed}. (Brukner and Zeilinger 2001:4, \small{my emphasis})
\end{quote} 

This argument is problematic, however. Let us begin with the obvious point that a failure of the argument for uniqueness does not automatically rule out the Shannon information as a measure of uncertainty. In fact, the Shannon information can be seen as one of a general class of measures of uncertainty, characterised by a set of axioms in which the grouping axiom does not appear (Uffink 1990), hence the grouping axiom is not necessary for the interpretation of the Shannon information as a measure of uncertainty. (Uffink in fact has previously argued that the grouping axiom is not a natural constraint on a measure of information and should not be imposed as a necessary constraint, even in the classical case (Uffink 1990, \S 1.6.3).) So from the fact that on the Brukner/Zeilinger reading, the grouping axiom seems to embody some classical assumptions that do not hold in the quantum case, it does not follow that the concept of the Shannon information as a measure of uncertainty involves those classical assumptions.

Furthermore, Brukner and Zeilinger's grouping axiom is not in fact equivalent to the standard form and the standard form is equally applicable in both the classical and quantum cases. Thus the Shannon information has not been shown to involve classical assumptions and the standard axiomatic derivation can indeed go through in the quantum context. The probabilities appearing in Grouping Axiom \ref{faddeevgroupingaxiom} are well defined in both the classical and quantum cases. 

\begin{figure}
\begin{center}
\includegraphics*[151,612][388,725]{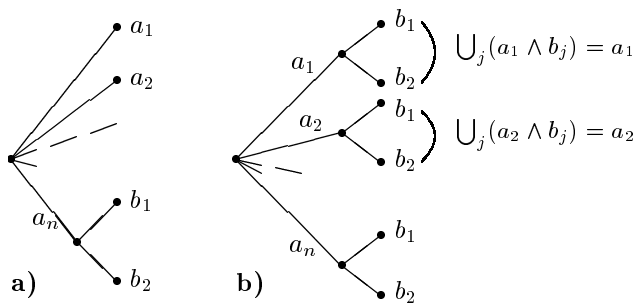}
\end{center}
\caption{{\bf a}) in Grouping Axiom \ref{faddeevgroupingaxiom} and eqn.~(\ref{conditional}), the `$B$' outcomes $b_{1}$ and $b_{2}$ cannot occur without $a_{n}\stackrel{\rm def}{=}b_{1}\cup b_{2}$ occurring. {\bf b}) in the joint experiment scenario, $b_{1}$ or $b_{2}$ can occur without $a_{n}$ occurring, but this is made to appear the same by coarse graining the joint experiment and only recording $B$ values when we get $A$ outcome $a_{n}$.}\label{difference}
\end{figure}

In Brukner and Zeilinger's notation, Grouping Axiom \ref{faddeevgroupingaxiom} would be written as
\begin{multline}\label{conditional}  
H(p(a_{1}),p(a_{2}),\ldots,p(a_{n-1}),p(b_{1}),p(b_{2}))     \\
 = H(p(a_{1}),\ldots,p(a_{n-1}),p(b_{1}\vee b_{2})) \\
 +p(b_{1}\vee b_{2})H(p(b_{1}|b_{1}\vee b_{2}),p(b_{2}|b_{1}\vee b_{2})) 
\end{multline}
This refers to an experiment with $n+1$ outcomes labelled by $a_{1},\ldots a_{n-1},b_{1},b_{2}$ and the grouping of two of these outcomes together, and is clearly different from Grouping Axiom \ref{BZ1} (see Fig. \ref{difference}).
Brukner and Zeilinger's Grouping Axiom \ref{BZ1} is the result of applying (\ref{conditional}) to a coarse grained joint experiment in which we only distinguish the $B$ outcomes of $A\wedge B$ for $A$ outcome $a_{n}$, and Grouping Axiom \ref{BZ2} is the result of applying the simple grouping axiom to the fine grained joint experiment $n$ times. These are not, then, expressions of the grouping axiom, but rather demonstrate its effect when actually applied to an already well defined joint probability distribution. The absence of certain joint probability distributions in quantum mechanics does not, however, affect the meaningfulness of the grouping axiom, because in its proper formulation it does not refer to joint experiments\footnote{To see that the standard case and the joint experiment case are mathematically distinct, note that the joint experiment formalism cannot express the situation in which $B$ events only happen if the $a_{n}$ event occurs. For that we would require that $p(a_{n})=p(b_{1})+p(b_{2})$, but then the marginal distribution for $B$ outcomes in the joint experiment does not sum to unity as is required for a well defined joint experiment, $\sum_{i} p(b_{i})=p(a_{n}) \neq 1$ in general.}
. (Note that if the outcomes of the experiment in eqn.~(\ref{conditional}) were represented by one dimensional projection operators, the event $b_{1}\vee b_{2}$ would be represented by a sum of orthogonal projectors which commutes with the remaining projectors; a similar relation (\textit{coexistence}) holds if the outcomes are represented by POV elements (\textit{effects}), Busch \textit{et al.} (1996).)

Thus we see that the explicit argument fails. There may remain, however, a certain intuitive one. Brukner and Zeilinger are perhaps suggesting that we miss something importantly quantum by using the Shannon information as its derivation is restricted to the case of commuting observables. Or rather (since we have seen that the grouping axiom does not explicitly refer to joint experiments and we know that it is not in fact necessary for the functioning of the Shannon information as a measure of uncertainty anyway), because it can only tell us about the uncertainty in joint experiments for mutually commuting observables rather than the full set of observables. However, when we recall that a measure of uncertainty is a measure of the spread of a probability distribution, we see that this simply amounts to the truism that one cannot have a measure of the spread of a joint probability distribution unless one actually has a joint probability distribution, and it in no way implies that there is anything un-quantum about the Shannon information itself.

The fact that joint probability distributions cannot be defined for all possible groupings of experiments that we might consider does not tell us anything about whether a certain quantity is a good or bad measure of uncertainty for probability distributions that \textit{can} be defined (e.g. any probability distribution derived from a quantum state by the trace rule, or conditional probabilities given by the L\"uders rule). We must be careful not to confuse the question of what makes a good measure of uncertainty with the question of when a joint probability distribution can be defined.

We already know that a function of a joint probability distribution cannot be a way of telling us how much we know, or how uncertain we are, \textit{in general} when we know the state of a system, because we know that a joint probability distribution for all possible measurements does not exist. It is for this reason in quantum mechanics that we introduce measures of mixedness such as the von Neumann entropy, which are functions of the \textit{state} rather than of a probability distribution. It is not a failing of the Shannon information as a measure of uncertainty or expected information gain that it does not fulfil the same role.
Part of Brukner and Zeilinger's worry about the Shannon information thus seems to arise because they are trying to treat it too much like a measure of mixedness, a measure of how uncertain we are in general when we know the state of a quantum system\footnote{This is illustrated for example in their reply to criticism of their grouping axiom argument by Hall (Brukner and Zeilinger 2000b, Hall 2000).
Hall presents an interpretation of the grouping axiom concerning the increase in randomness on mixing of non-overlapping distributions, to which Brukner and Zeilinger's worries about joint experiments would not apply. Their reply, in essence, is that the density matrix cannot be simultaneously diagonal in non-commuting bases, therefore it cannot be thought to be composed of non-overlapping classical distributions, hence Hall's grouping axiom will not apply, further supporting their original claim that the Shannon measure is tied to the notion of classical properties. What this reply in fact establishes, however, is that Hall's axiom applied to mixtures of \textit{classical} distributions is not relevant to characterising the randomness of the \textit{density matrix}; but this is something with which everyone would agree, and this job certainly not one for which the Shannon information is intended. (If we did wish to use the grouping axiom in characterising the randomness of the density matrix, we would apply Hall's version to mixtures of density operators with orthogonal support; this would then pick out the von Neumann entropy up to a constant factor (Wehrl 1978).)}. 

\section{Brukner and Zeilinger's `Total information content'\label{finalargument}}

The final argument proposed against the Shannon information is that it is not appropriately related to a notion of `total information content' for a quantum system. It is also suggested that the von Neumann entropy, which has a natural relation to the Shannon information, is not a measure of information content as it makes no explicit reference to information gain from measurements in general (Brukner and Zeilinger 2001, 2000b).

In place of the Shannon information, Brukner and Zeilinger propose the quantity
\begin{equation}
I(\vec{p})={\cal N}\sum_{i=1}^{n}\left (p_{i}-\frac{1}{n}\right )^{2}, \label{BZ info}
\end{equation}
from which they derive their notion of total information content as follows:

\begin{sloppypar}
A set of measurements is called \textit{mutually unbiased} or \textit{complementary} (Schwinger 1960) if the sets of projectors $\{P\}, \{Q\}$ associated with any pair of measurement bases satisfy $\mbox{Tr}(PQ)=1/n$, where $n$ is the dimensionality of the system. There can exist at most $n+1$ such bases (Wootters and Fields 1989), constituting a \textit{complete} set,
and as was shown by Ivanovic (1981), measurement of such a complete set on an ensemble of similarly prepared systems determines their density matrix $\rho$ completely. In analogy to acquiring the information associated with a (pointlike) classical system by learning its state (determining its position in phase space), Brukner and Zeilinger then suggest that the \textit{total information content} of a quantum system should be given by a sum of information measures for a complete set of mutually unbiased measurements. Taking $I(\vec{p})$ as the measure of information, the result is  a unitarily invariant quantity:
\begin{equation}\label{totalinformation}
I_{tot}=\sum_{j=1}^{n+1}I(\vec{p^{j}})=\sum_{ji}\left( p_{i}^{j}-\frac{1}{n} \right) ^{2}=\mbox{Tr}\left( \rho -\frac{\mathbf{1}}{n}\right)  ^2.
\end{equation}

This also provides their argument against the Shannon information. It is a necessary constraint on a measure of total information content, they argue, that it be unitarily invariant, but substituting $H(\vec{p})$ for $I(\vec{p})$ in (\ref{totalinformation}) does not result in a unitarily invariant quantity, that is, we do not have a sum to a `total information content'. Let us call the requirement that a measure sum to a unitarily invariant quantity for a complete set of mutually unbiased measurements the `total information constraint'. The suggestion is that the Shannon measure is inadequate as a measure of information gain because it does not satisfy the total information constraint and hence does not tell us how much of the total information content of a system we learn by performing measurements in a given basis. Similarly, the complaint against the von Neumann entropy is that it is merely a measure of mixedness, as unlike $I_{tot}$, it has no relation to the information gained in a measurement unless we happen to measure in the eigenbasis of $\rho$.
\end{sloppypar}

A few remarks are in order. First, $I(\vec{p})$ and $I_{tot}$ are not unfamiliar expressions. The quantity $\sum_{i}(p_{i}-1/n)^2$ is one of the class of measures of the concentration of a probability distribution given by Uffink (1990), and Fano (1957), for example, remarks that $\mbox{Tr}(\rho^2)$ can serve as a good measure of information; furthermore, the relation expressed in eqn.~(\ref{totalinformation}) has previously been employed by Larsen (1990) in discussing exact uncertainty relations. Note also that $I(\vec{p})$ is an \textit{increasing} function of the concentration of a probability distribution, hence a measure of \textit{how much we know} given a probability distribution, rather than being a measure of uncertainty like $H(\vec{p})$; similarly $I_{tot}$ is an \textit{increasing} function of the purity of $\rho$.

More importantly, however, `information content' might mean several different things. It may not, then, be reasonable to require that every meaningful information measure sum to a unitarily invariant quantity that can be interpreted as an information content. Moreover, we may well ask why information measures for a complete set of mutually unbiased measurements should be expected to sum to \textit{any} particularly interesting quantity.
That the measure $I(\vec{p})$ happens to sum to a unitarily invariant quantity is, as we shall presently see, the consequence of a geometric property tangential to its role as a measure of information.

\subsection{Some Different Notions of Information Content}\label{differentnotions}

It is useful to distinguish between the information encoded in a system, the information required to specify the state of a system (more precisely, the information required to specify a \textit{sequence} of states drawn from a given ensemble) and states of complete and less than complete knowledge or information. Each of these can serve as a notion of information content in an appropriate context. 
In the classical case, their differences can be largely ignored, but in the quantum case there are important divergences. It is, for instance, necessary to introduce the concept of the accessible information to characterise the difference between information encoded and specification information (Schumacher 1995).

If we consider encoding the outputs $a_{i}$ of a classical information source $A$ into pure states $|a_{i}\rangle$ of an ensemble of quantum systems, then the state of the ensemble will be given by $\rho=\sum_{i}p(a_{i})|a_{i}\rangle\langle a_{i} |$. The von Neumann entropy, $S(\rho)=-\mbox{Tr}\rho\log\rho$, is a measure of how mixed this state is, giving us one sense of information content - the more mixed a state, the less information we have about what the outcome of measurements on systems described by the state will be\footnote{Mixed states are also sometimes said to be states of less than complete information due to a lack of information about the way a system was prepared, represented by a probability distibution over possible pure states. Our reading is to be preferred given the many-one relation of preparation procedures to density operators and the fact that density operators can also result from tracing out unwanted degrees of freedom.}.

If we are presented with a sequence of systems drawn from an ensemble prepared in this manner, each will be in one of the pure states, and the number of bits per system required to specify this sequence will be $H(A)$, the information of the classical source (which will be greater than $S(\rho)$ unless we have coded in orthogonal states). This is the \textit{specification information}, also the amount of information required to prepare the sequence. For the amount of information that has actually been encoded into the systems, however, we need to consider measurements on the ensemble and the Shannon mutual information $H(A:B)$.

As already remarked (Section~\ref{preexistingbits}), if we encode using a certain basis (our $|a_{i}\rangle$ form an orthonormal set) and we measure in a different basis, then $H(A:B)< H(A)$; quantum `noise' reduces the amount we can decode. More significantly, if we have coded in non-orthogonal states (if, for example the number of outputs of our classical source is greater than the dimensionality of our quantum systems), then no measurement can distinguish these states perfectly and we cannot recover the complete classical information.   
Taking `encoded' to be a `success' word  (something cannot be said to have been encoded if it cannot in principle be decoded), then the maximum amount of information encoded in a system is given by the \textit{accessible information}, the maximum over all decoding observables of the mutual information. Using non-orthogonal coding states, the amount we can encode is less than the specification information. A well known result due to Holevo (1973) provides an upper bound on the mutual information resulting from measurement of any observable (including POV measurements). For the general case of encoding states $\rho_{a_{i}}$, this is given by
\[H(A:B) \leq S(\rho) - \sum_{i} p(a_{i})S(\rho_{a_{i}}),\] 
which in the case we are considering of pure encoding states, reduces to $H(A:B)\leq S(\rho)$. This provides a very strong sense in which the von Neumann entropy does give us a notion of the total information content of a quantum system --- it is the maximum amount that can actually be encoded in the system.      

Brukner and Zeilinger do not consider a quantum communication channel but are concerned rather with the information content of a single system considered in isolation. 
This information content is supposed to relate to how much we learn from learning the state, but if the system is being treated in isolation then by learning its state we are not acquiring a certain amount of information in virtue of the state being drawn from a given ensemble, as in the standard notion of information.
(Hence their analogy with gaining the information content of a classical system fails to hold.) 
In fact, their `total information content' seems best interpreted as a measure of mixedness analogous to the von Neumann entropy.

When introduced (Brukner and Zeilinger 1999), the information measure $I(\vec{p})$ is presented as a measure of how much we know about what the outcome of a particular experiment will be, given the state. The \textit{total} information of the state, then, would seem to be a measure of how much we know \textit{in general} about what the outcomes of experiments will be given the state; and this is precisely a question of the degree of mixedness of the state\footnote{Recently it has been noted that $I_{tot}$ is also related to the average distance of our estimate of the unknown state from the true state (measured in the Hilbert-Schmidt norm), given only a finite number $N$ of experiments in each mutually unbiased basis (Rehacek and Hradil 2002). This seems best understood as indicating that the mean error in our state estimation is inversely related to $N$, with a constant of proportionality that depends on the dimension of the system and the mixedness of the state. In any case, Brukner and Zeilinger are primarily interested in how much we know when the state has been determined to arbitrary accuracy.}. 

\subsubsection{Measures of mixedness}\label{mixedness}

The functioning of measures of mixedness can usefully be approached via the notions of majorization and Schur convexity (concavity).
The majorization relation $\prec$ imposes a pre-order on probability distributions (Uffink 1990, Nielsen 2001). A probability distribution $\vec{q}$ is majorized by $\vec{p}$, $\vec{q}\prec\vec{p}$, iff $q_{i}=\sum_{j}S_{ij}p_{j}$, where $S_{ij}$ is a doubly stochastic matrix. That is (via Birkhoff's theorem), if $\vec{q}$ is a mixture of permutations of $\vec{p}$. Thus if $\vec{q} \prec \vec{p}$, then $\vec{q}$ is a more mixed or disordered distribution than $\vec{p}$.

Schur convex (concave) functions respect the ordering of the majorization relation: a function $f$ is Schur convex if, if $\vec{q}\prec \vec{p}$ then $f(\vec{q})\leq f(\vec{p})$, and Schur concave if, if $\vec{q}\prec \vec{p}$ then $f(\vec{q})\geq f(\vec{p})$ (for strictly Schur convex(cave) functions, equality holds only if $\vec{q}$ and $\vec{p}$ are permutations of one another). This explains the utility of such functions as measures of the concentration and uncertainty of probability distributions, respectively.

The majorization relation will apply equally to the vectors of eigenvalues of density matrices. 
It can be shown that the vector of eigenvalues $\vec{\lambda^{\prime}}$ of the density matrix $\rho^{\prime}$ of the post measurement ensemble for a (non-selective) projective measurement is majorized by the vector of eigenvalues $\vec{\lambda}$ of the pre-measurement state $\rho$ (Nielsen 2001). (If we measure in the eigenbasis of $\rho$, then there is, of course, no change in the eigenvalues). The $\lambda^{\prime}_{i}$ are just the probabilities of the different outcomes of the measurement in question, thus the probability distribution for the outcomes of any given measurement will be more disordered or spread than the eigenvalues of $\rho$.

If we take any Schur concave function we know to be a measure of uncertainty, for instance the Shannon information $H(\vec{p})$, and $\vec{p}$ is the probability distribution for measurement outcomes, we then know that $H(\vec{p})\geq H(\vec{\lambda})$, for any projective measurement we might perform. This explains why $H(\vec{\lambda})=S(\rho)$, a measure of mixedness, is a measure of how much we know given the state: the more mixed a state, the more uncertain we must be about the outcome of any given measurement. Similarly, if we take a measure of the concentration of a probability distribution, a Schur convex function such as $I(\vec{p})$, then we know that for any measurement with outcome probability distribution $\vec{p}$, $I(\vec{p})\leq I(\vec{\lambda})=I_{tot}$; and this explains why $I_{tot}$ is a measure of how much we know given $\rho$: the less the value of $I_{tot}$, the less able we are to predict the outcome of any given experiment.    

Brukner and Zeilinger would of course deny that their total information content is merely a measure of mixedness. The argument that it is more than this rests on the satisfaction of the total information constraint, the relation between the measure of information $I(\vec{p})$ and $I_{tot}$ for a complete set of mutually unbiased measurements as expressed in eqn.~(\ref{totalinformation}). We shall now see that this relation can be given a simple geometric explanation using the Hilbert-Schmidt representation of density operators.

\subsection{The Relation between Total Information Content and $I(\vec{p})$}

The set of complex $n\times n$ Hermitian matrices forms an $n^{2}$-dimensional real Hilbert space $V_{h}(C^n)$ on which we have defined an inner product $(A,B)=\mbox{Tr}(AB); A,B \in V_{h}(C^n)$ and a norm $\|A\|=\sqrt{\mbox{Tr}(A^2)}$ (Fano 1957; Wichmann 1963).
The density matrix $\rho$ of an $n$ dimensional quantum system can be represented as a vector in this space. The requirements on $\rho$ of unit trace and positivity imply that the tip of any such vector must lie in the $n^2 -1$ dimensional hyperplane $\bm{T}$ a distance $1/\sqrt{n}$ from the origin and perpendicular to the unit operator $\mathbf{1}$, and on or within a hypersphere of radius one centred on the origin.

It is useful to introduce a set of basis operators on our space; we require $n^{2}$ linearly independent operators $U_{i}\in V_{h}(C^n)$ and it may be useful to require orthogonality: $\mbox{Tr}(U_{i}U_{j})=\mbox{const.}\times \delta_{ij}$.
Any operator on the system can then be expanded in terms of this basis and in particular, $\rho$ can be written as 
\[ \rho=\frac{\mathbf{1}}{n} + \sum_{i=1}^{n^{2}-1}\mbox{Tr}(\rho U_{i})U_{i},\]
where we have chosen $U_{0}=\mathbf{1}$ to take care of the trace condition.

Evidently, $\rho$ may be determined experimentally by finding the expectation values of the $n^{2}-1$ operators $U_{i}$ in the state $\rho$. If we include the operator $\mathbf{1}$ in our basis set, then the idempotent projectors associated with measurement of any maximal (non-degenerate) observable will provide a maximum of a further $n-1$ linearly independent operators. Obtaining the probability distribution for a given maximal observable will thus provide $n-1$ of the parameters required to determine the state, and the minimum number of measurements of maximal observables that will be needed in total is $n+1$, if each observable provides a full complement of linearly independent projectors. 

Each such set of projectors spans an $n-1$ dimensional hyperplane in $V_{h}(C^n)$ and their expectation values specify the projection of the state $\rho$ into this hyperplane. Ivanovic (1981) noted that projectors $P,Q$ belonging to any two different mutually unbiased bases will be orthogonal in $\bm{T}$, hence the hyperplanes associated with measurement of mutually unbiased observables are orthogonal in the space in which density operators are constrained to lie in virtue of the trace condition. If $n+1$ mutually unbiased observables can be found, then, $V_{h}(C^n)$ can be decomposed into orthogonal subspaces given by the one dimensional subspace spanned by $\mathbf{1}$ and the $n+1$ subspaces associated with the mutually unbiased observables. The state $\rho$ can then be expressed as:
\begin{equation}
\rho= \frac{\mathbf{1}}{n} +\sum_{j=1}^{n+1}\sum_{i=1}^{n}q_{i}^{j}\bar{P}_{i}^{j}, \label{rho}
\end{equation}
where $\bar{P}_{i}^{j}=P_{i}^{j}-\mathbf{1}/n$ is the projection onto $\bm{T}$ of the $i$th idempotent projector in the $j$th mutually unbiased basis set, and $q_{i}^{j}=(p_{i}^{j}-1/n)$ is the expectation value of this operator in the state $\rho$.
 For a given value of $j$, the vectors $\bar{P}_{i}$ span an $(n-1)$ dimensional orthogonal subspace and the square of the length of a vector expressed in the form (\ref{rho}) lying in subspace $j$ will be given by $\sum_{i=1}^{n}(q_{i}^{j})^2 = I(\vec{p^{j}})$.       

The geometrical explanation of $I_{tot}$ is then simply as follows. $\mbox{Tr}(\rho^{2})$ is the square of the length of $\rho$ in $V_{h}(C^n)$ and $\mbox{Tr}(\rho - \mathbf{1}/n)^{2}$ is the square of the distance of $\rho$ from the maximally mixed state (the length squared of $\rho$ in $\bm{T}$). This squared length will just be the sum of the squares of the lengths of the components of the vector $\rho-\mathbf{1}/n$ in the orthogonal subspaces into which we have decomposed $V_{h}(C^n)$, i.e. it will be given by $\sum_{ji}(q_{i}^{j})^2$. This is what eqn.~(\ref{totalinformation}) reports and it explains how $I_{tot}$ and $I(\vec{p})$ satisfy the total information constraint.

Thus we see that if $I_{tot}$ differs from being a simple measure of mixedness, then that is because it is a measure of length also; and this explains why it can be given by a sum of quantities $I(\vec{p})$ for a complete set of mutually unbiased measurements. As measures of \textit{how much we know given the state}, however, $I_{tot}$ and $S(\rho)$ bear the same relation to their appropriate measures of information, as we saw in the previous section. Equally, as measures of information, $H(\vec{p})$ stands to $S(\rho)$ in the same relation as $I(\vec{p})$ to $I_{tot}$.  
$I_{tot}$ is the \textit{upper} bound on the amount we can \textit{know} about the outcome of a measurement as measured by $I(\vec{p})$; $S(\rho)$ is the \textit{lower} bound on our \textit{uncertainty} about what the outcome of a measurement will be, as measured by $H(\vec{p})$.

The complaint against the Shannon information was supposed to be that as $H(\vec{p})$ fails to satisfy the total information constraint, it does not tell us the information gained from a particular measurement; the complaint against the von Neumann entropy that as $S(\rho)$ is not given by a sum of measures for a complete set of mutually unbiased measurements, it is not suitably related to the information gained from a measurement.
However, we can now see that insisting on the total information constraint in this way is tantamount to insisting that only a function which measures the length of the component of $\rho$ lying in a given hyperplane can be a measure of information, and correlatively, that the only viable notion of total information content is a measure of the length of $\rho$ in $V_{h}(C^{n})$.
But $H(\vec{p})$ can be a perfectly good measure of information without having to be a measure of the length of the projection of $\rho$ into the subspace associated with an observable; and as we have just seen, $S(\rho)$ does have an explicit relation to the information gain from measurement that justifies its interpretation as a total information content. A relation, moreover, that $I_{tot}$ also possesses and which serves to justify \textit{its} interpretation as a measure of how much we know given the state.

Hence our conclusion must be that the total information constraint is not a reasonable requirement on measures of information.   

Of course, $H(\vec{p})$ does not tell us the information gain on measurement it we take, as Brukner and Zeilinger seem to, `the information encoded in a basis' simply to \textit{mean} the length squared of the component of the state lying in the measurement hyperplane; but this is a non-standard usage. $H(\vec{p})$ certainly remains a measure of our expected information gain from performing a particular measurement (how much the outcome will surprise us, on average, given that we have the probability distribution); and if we are interested in the amount of information encoded, in the usual sense of the word, that can be decoded using a particular measurement, i.e. if we have a string of systems into which information has actually been encoded, then we may always just consider the Shannon mutual information associated with that measurement. (The `total information' associated with this quantity will then be given, via the Holevo bound, as the von Neumann entropy.)

\section{Conclusion}

Of the three arguments that Brukner and Zeilinger have presented against the Shannon information, we have seen that the first two fail outright. These arguments sought to establish that the notion of the Shannon information is undermined in the quantum context due to a reliance on classical concepts. With regard to the first we saw that, contrary to Brukner and Zeilinger, the existence of a pre-determined string is neither necessary nor sufficient for the interpretation of $H(\vec{p})$ as a measure of information, hence the absence of such a string would not cause any problems for the Shannon information in quantum mechanics.

The objective of their second argument was to highlight classical assumptions in the grouping axiom that would prevent the axiomatic derivation of the Shannon information going through in the quantum case. This argument turned out to be based on an erroneous reading of the grouping axiom that appeals to joint experiments. The grouping axiom is in fact perfectly well defined in the quantum case and the standard axiomatic derivation of the form of $H(\vec{p})$ can indeed go through. 
The grouping axiom does not reveal any problematic classical assumptions implicit in the Shannon information.

In their final argument, Brukner and Zeilinger suggest that defining the notion of the total information content of a quantum system in terms of the Shannon information would lead to a quantity with the unnatural property of unitary non-invariance. But this is not a compelling argument against the Shannon quantity as a measure of information. We have seen that it is not a necessary requirement on every meaningful measure of information that it sum to a unitarily invariant quantity for a complete set of mutually unbiased measurements; nor, conversely, is it necessary that every viable notion of total information content be given by such a sum of individual measures of information. 

Brukner and Zeilinger's arguments thus fail to establish that the Shannon information involves any particularly classical assumptions or that there is any difficulty in the application of the Shannon measure to measurements on quantum systems. The Shannon information is perfectly well defined and appropriate as a measure of information in the quantum context as well as in the classical.

\newpage

\appendix
\section{Appendix: Zeilinger's Foundational Principle}

As remarked in the introduction, Brukner and Zeilinger's discussion of the Shannon information follows on from Zeilinger's (1999) proposal of an information-theoretic foundational principle for quantum mechanics. This principle, he suggests, is to play a role in quantum mechanics similar to that of the Principle of Relativity in Special Relativity, or to the Principle of Equivalence in General Relativity. Like these, the Foundational Principle is to be an intuitively understandable principle which plays a key role in deriving the structure of the theory. In particular, Zeilinger suggests that the Foundational Principle provides an explanation for the irreducible randomness in quantum measurement and for the phenomenon of entanglement. For discussion of how the Foundational Principle motivates the arguments against the Shannon information, see Timpson (2002); in this Appendix we shall consider whether the Principle can indeed be successful as a foundational principle for quantum mechanics. 

Before stating the Foundational Principle, it is helpful to identify two philosophical assumptions that  Zeilinger's position incorporates. The first is a form of phenomenalism: physical objects are taken not to exist in and of themselves, but to be mere constructs relating sense impressions (Zeilinger 1999:633)\footnote{Here I take phenomenalism to be the doctrine that the subject matter of all conceivable propositions are one's own actual or possible experiences, or the actual and possible experiences of another.}; the second assumption is an explicit instrumentalism about the quantum state:
\begin{quote}
The initial state...represents all our information as obtained by earlier observation...[the time evolved] state is just a short-hand way of representing the outcomes of all possible future observations. (Zeilinger 1999:634)
\end{quote}
With these assumptions noted, let us consider the two distinct formulations of the Principle presented in Zeilinger (1999):
\newtheorem{FP}{FP}
\begin{FP}\label{FP1}
An elementary system represents the truth value of one proposition.
\end{FP}
\begin{FP}\label{FP2}
An elementary system carries one bit of information.
\end{FP}

At first glance, these two statements appear most naturally to be concerned with the amount of information that can be encoded into a physical system. However, this interpretation is at odds with the passage in which Zeilinger motivates the Foundational Principle. In this passage, his concern is with the number of propositions required to describe a system. He considers the analysis of a composite system into constituent parts and remarks that it is natural to assume that each constituent system will require fewer propositions for its description than the composite does. The end point of the analysis will be reached when we have systems described by a single proposition only; and it is these systems that are termed `elementary'. 

The situation is clarified when Zeilinger goes on to explain what he means by an elementary system carrying or representing some information:
\begin{quote}
...that a system ``represents" the truth value of a proposition or that it ``carries" one bit of information only implies a statement concerning what can be said about possible measurement results. (Zeilinger 1999:635)
\end{quote}
Thus the Foundational Principle is not a constraint on how much information can be encoded into a physical system. It is a constraint on how much the state of an elementary system can say about the results of measurement. This interpretation is rendered consistent with the discussion in terms of the propositions required to describe a system, as from Zeilinger's instrumentalist point of view, describing (the state of) a quantum system can only be to make a claim about future possible measurement results. Furthermore, we can understand the peculiar idiom of a system `representing' some information, where this is taken not to refer to the encoding of some information into a system, when we recall that from the point of view of Zeilinger's phenomenalism, a physical system is not an actual thing. On his view, a system represents a quantity of information about measurement results because a physical system literally \textit{is} nothing more than an agglomeration of actual and possible sense impressions arising from observations.

In short, however, it seems that a clearer, and perhaps more philosophically neutral, statement of the Foundational Principle would be the following:
\begin{FP}\label{FP3}
The state of an elementary system specifies the answer to a single yes/no experimental question,
\end{FP}
where we have used the fact that by `proposition' Zeilinger means something that represents an experimental question. With this relatively clear statement of the Foundational Principle in hand, let us now consider its claims as a foundational principle for quantum mechanics. 

To begin with, we should note the limitations implied by Zeilinger's conception of the description of a system. It might not always be the case that the state of an individual system can be characterised appropriately as a list of experimental questions to which answers are specified; and in such a case, the terms of the Foundational Principle cannot be set up. Consider the de Broglie-Bohm theory, for example, with its elements of holism and contextuality --- even though the theory is deterministic, the results of measurements are in general not determined by the properties of the object system alone but are the result of interaction between object system and measuring device. It would seem that this theory could neither be supported nor ruled out by the Foundational Principle, as we can neither identify something that would count as an elementary system in this theory, given the way `elementary system' has been defined, nor, \textit{a fortiori}, begin to enumerate how many experimental questions such an entity might specify. However, for present purposes, let us put this sort of worry to one side. 

Another concern arises when considering the distinction we have drawn between describing a system and encoding information into it. Unlike encoding, the notion of describing a system presupposes a certain language in which the description is made, and the description of a given system could be longer or shorter depending on the conceptual resources of the language used. If we are to make a claim about the number of propositions required to describe a system, then, as we must when identifying an elementary system to figure in the Foundational Principle, we must already have made a choice of the set of concepts with which to describe the system. But this is worrying if the purpose of the Foundational Principle is to serve as a basis from which the structure of our theory is to be derived. If we already have to make substantial assumptions about the correct terms in which the objects of the theory are to be described, then it may be that the Foundational Principle will be debarred from serving its foundational purpose.
With this worry in mind, let us now consider the first of the concrete claims for the Foundational Principle, that it explains the irreducible randomness of quantum measurements. 

Zeilinger's suggestion is that we have randomness in quantum mechanics because:
\begin{quote}
...an elementary system cannot carry enough information to provide definite answers to all questions that could be asked experimentally (Zeilinger 1999:636),
\end{quote} 
and this randomness must be irreducible, because if it were reduced to hidden properties, then the system would carry more than one bit of information. Unfortunately, this does not constitute an explanation of randomness, even if we have granted the existence of elementary systems and adopted the Foundational Principle. For the following question remains: why is it that experimental questions exist whose outcome is not already determined by a specification of the finest grained state description we can offer? How is it that any space for randomness remains? Or again, why isn't one bit enough?
The point is, it has not been explained \textit{why} the state of an elementary system cannot specify an answer to all experimental questions: this does not in fact appear to follow from the Foundational Principle. 
The Foundational Principle says nothing about the structure of the set of experimental questions, yet this turns out to be all-important.

Consider the case of a classical Ising model spin, which has only two possible states, `up' or `down'; here one bit, the specification of an answer to a single experimental question (`Is it up?') \textit{is} enough to specify an answer to all questions that could be asked. There is no space for randomness here, yet this classical case is perfectly consistent with the Foundational Principle. Thus it seems that no explanation of randomness is forthcoming from the Foundational Principle and furthermore, it is far from clear that the Principle, on its own, in fact allows us to distinguish between quantum and classical. 

Of course, if one assumes that experimental questions are represented in the quantum way, as projectors on a Hilbert space, then even for the simplest non-trivial state space, there will be non-equivalent experimental questions, the answer to one of which will not provide an answer to another; but we cannot assume this structure if it is the very structure that we are trying to derive. It appears from the way in which the Foundational Principle is supposed to be functioning in the attempted explanation of randomness, that something like the quantum structure of propositions is being assumed. But this is clearly fatal to the prospects of the Foundational Principle as a foundational principle.\footnote{In a sense, we could say that Zeilinger's explanation of randomness is problematic because it fails to explain why the state space of quantum mechanics is so gratuitously large from the point of view of storing information (Caves and Fuchs 1996). It is then striking that this attempted information-theoretic foundational approach to quantum mechanics has not allowed for one of the significant insights vouchsafed by quantum information theory.} 

Does the Principle fare any better with the proposed explanation of entanglement? The idea here is to consider $N$ elementary systems, which, following from the Foundational Principle, will have $N$ bits of information associated with them. The suggestion is that entanglement results when all $N$ bits are exhausted in specifying joint properties of the system, leaving none for individual subsystems (Zeilinger 1999), or more generally, when more information is used up in specifying joint properties than would be possible classically. The underlying thought is that this approach captures the intuitive idea that when we have an entangled system, we know more about the joint system (which may be in a pure state) than we do about the individual sub-systems (which must be mixed states). 
The proposal is further developed in (Brukner \textit{et al} 2001), where Brukner and Zeilinger's information measure is used to provide a quantitative condition for $N$ qubits to be unentangled, which is then related to a condition for the violation of a certain $N$-party Bell inequality. 

To give a basic example of how the idea is supposed to work, consider the case of two qubits. Recall that the maximally entangled Bell states are joint eigenstates of the observables $\sigma_{x}\otimes\sigma_{x}$ and $\sigma_{y}\otimes\sigma_{y}$. From the Foundational Principle, only two bits of information are associated with our two systems, i.e. the states of these systems can specify the answer to two experimental questions only. If the two questions whose answers are specified are `Are both spins in the same direction along $x$?' ($1/2(\mathbf{1}\otimes\mathbf{1} + \sigma_{x}\otimes\sigma_{x})$) and `Are both spins in the same direction along $y$?' ($1/2(\mathbf{1}\otimes\mathbf{1} + \sigma_{y}\otimes\sigma_{y})$), then we end up with a maximally entangled state. If, by contrast, the two questions had been `Are both spins in the same direction along $x$?' and `Is the spin of particle 1 up along $x$?', the information would not have all been used up specifying \textit{joint} properties and we would have instead a product state (joint eigenstate of $\sigma_{x}\otimes\sigma_{x}$ and $\sigma_{x}\otimes \mathbf{1}$).    

Now, although this idea may have its attractions when used as a criterion for entanglement within quantum mechanics, it does not succeed in providing an explanation for the phenomenon of entanglement, which was the original claim.

If we return to the starting point and consider our $N$ elementary systems, all that the Foundational Principle tells us  regarding these systems is that their individual states specify the answer to a single yes/no question concerning each system individually. There is, as yet, no suggestion of how this relates to joint properties of the combined system. Some assumption needs to be made before we can go further. For instance, we need to enquire whether there are supposed to be experimental questions regarding the joint system which can be posed and answered that are not equivalent to questions and answers for the systems taken individually. (We know that this will be the case, given the structure of quantum mechanics, but we are not allowed to \textit{assume} this structure, if we are engaged in a foundational project.\footnote{To illustrate, a simultaneous truth value assignment for the experiments $\sigma_{x}\otimes\sigma_{x}$ and $\sigma_{y}\otimes\sigma_{y}$ cannot be reduced to one for experiments of the form $\mathbf{1}\otimes\mathbf{a}.\boldsymbol{\sigma}, \mathbf{b}.\boldsymbol{\sigma}\otimes\mathbf{1}$.}) If this \textit{is} the case then there can be a difference in the information associated with correlations (i.e., regarding answers to questions about joint properties) and the information regarding individual properties. But then we need to ask: why is it that there exist sets of experimental questions to which the assignment of truth values is not equivalent to an assignment of truth values to experimental questions regarding individual systems? 

Because such sets of questions exist, more information can be `in the correlations' than in individual properties. Stating that there is more information in correlations than in individual properties is then to report that such sets of non-equivalent questions exist, \textit{but it does not explain why they do so}. However, it is surely this that demands explanation --- why is it not simply the case that all truth value assignments to experimental questions are reducible to truth value assignments to experimental questions regarding individual properties, as they are in the classical case? That is, why does entanglement exist? In the absence of an answer to the question when posed in this manner, the suggested explanation following from the Foundational Principle seems dangerously close to the vacuous claim that entanglement results when the quantum state of the joint system is not a separable state. 

Of course, if we are in the business of looking within quantum mechanics and asking how product and entangled states differ, then it is indeed legitimate to consider something like the condition Brukner \textit{et al.} (2001) propose; and we can then consider how good this condition is as a criterion for entanglement\footnote{At this point it is worth noting that there have been other discussions of entanglement which develop the intuitive idea that when faced with entangled states, we know more about joint properties than individual properties. A very general framework is presented by Nielsen and Kempe (2001), who use the majorization relation to compare the spectra of the global and reduced states of the system; a necessary condition for a state to be separable is then that it be more disordered globally than locally.}. But as mentioned before, if we are trying to explain the existence of entanglement then we cannot simply assume the quantum mechanical structure of experimental questions. 

Let us close by considering a final striking passage. Zeilinger suggests that the Foundational Principle might provide an answer to Wheeler's question `Why the quantum?' (Wheeler 1990) in a way congenial to the Bohrian intuition that the structure of quantum theory is a consequence of limitations on what can be said about the world:
\begin{quote}
The most fundamental viewpoint here is that the quantum is a consequence of what can be said about the world. Since what can be said has to be expressed in propositions and since the most elementary statement is a single proposition, quantization follows if the most elementary system represents just a single proposition. (Zeilinger 1999:642)
\end{quote}
But this passage contains a crucial non-sequitur. Quantization only follows if the propositions are projection operators on a complex Hilbert space. And why is it that the world has to be described that way? \textit{That} is the question that would need to be answered in answering Wheeler's question; and it is a question which, I suggest, the Foundational Principle goes no way towards answering.

\section*{Acknowledgements}
Thanks are due to Harvey Brown for useful discussion and in particular, for emphasis of the point raised in footnote~(\ref{KS}); thanks also to Jos Uffink and Michael Hall for comments. This work was supported by a studentship from the UK Arts and Humanities Research Board.

\section*{References}

Ash, R. (1965). Information Theory. New York London Sydney:Interscience Publishers.

Bell, J.S. (1982). On the Impossible Pilot Wave. Foundations of Physics 12, 989-999.

Brukner C. and Zeilinger A. (1999). Operationally Invariant Information in Quantum Measurements. Physical Review Letters 83(17), 3354. 

Brukner C. and Zeilinger, A. (2000a). Encoding and decoding in complementary bases with quantum gates. Journal of Modern Optics 47(12), 2233-2246.  

Brukner C. and Zeilinger, A. (2000b). Quantum Measurement and Shannon Information, A Reply to M.J.W. Hall. arXiv:quant-ph/0008091 

Brukner C. and Zeilinger, A. (2001). Conceptual inadequacy of the Shannon information in quantum measurements. Physical Review A 63, 022113.

Brukner C., Zukowski, M. and Zeilinger, A. (2001). The Essence of Entanglement. arXiv:quant-ph/0106119.

Bub, J. (1997). Interpreting the Quantum World. Cambridge University Press.

Busch, P., Lahti, P.J. and Mittelstaedt, P. (1996). The Quantum Theory of Measurement. Berlin Heidelberg: Springer-Verlag.

Caves, C.M and Fuchs, C.A. (1996). Quantum Information: How Much Information in a State Vector? In A. Mann and R. Revzen (Eds.), The Dilemma of Einstein, Podolsky and Rosen - 60 Years Later. Israel Physical Society. arXiv:quant-ph/9601025.

Faddeev, D.K. (1957). In H. Grell (Ed.), Arbeiten zum Informationstheorie I (pp.88-91). Berlin: Deutscher Verlag der Wissenchaften.

Fano, U. (1957). Description of States in Quantum Mechanics by Density Operator Techniques. Reviews of Modern Physics 29(1), 74-93.

Fuchs, C.A. (2000). Notes on a Paulian Idea. arXiv:quant-ph/0105039.

Fuchs, C.A. (2002). Quantum Mechanics as Quantum Information (and only a little more). arXiv:quant-ph/0205039.

Hall, M.J.W. (2000). Comment on `Conceptual Inadequacy of Shannon Information...' by C. Brukner and A. Zeilinger. arXiv:quant-ph/0007116.

Holevo, A.S. (1973). Information theoretical aspects of quantum measurement. Problems of Information Transmission (USSR) 9, 177-183.

Ivanovic, I.D. (1981). Geometrical Description of Quantal State Determination. Journal of Physics A 14, 3241-3245.

Jaynes, E.T. (1957). Information Theory and Statistical Mechanics. Physical Review 106(4), 620-630.

Larsen, U. (1990). Superspace Geometry: the exact uncertainty relationship between complementary aspects. Journal of Physics A 23, 1041-1061.

Maassen, H and Uffink, J.B.M. (1988). Generalized Entropic Uncertainty Relations. Physical Review Letters 60(12), 1103-1106.

Mermin, N.D. (2001). Whose Knowledge? In R. Bertlmann and A. Zeilinger (Eds.), Quantum Unspeakables: Essays in Commemoration of John S. Bell. Berlin Heidelberg: Springer-Verlag, forthcoming.

Nielsen, M.A. (2001). Characterizing mixing and measurement in quantum mechanics. Physical Review A 63, 022114.

Nielsen, M.A. and Kempe, J. (2001). Separable states are more disordered globally than locally. Physical Review Letters 86, 5184-5187.

Peres, A. (1995). Quantum Theory: Concepts and Methods. Dordrecht Boston London: Kluwer Academic Publishers.

Rehacek, J and Hradil, Z. (2002). Invariant Information and Quantum State Estimation. Physical Review Letters 88, 130401.

Schumacher, B. (1995). Quantum Coding. Physical Review A 51(4), 2738.

Schwinger, J (1960). Unitary Operator Bases. Proceedings of the National Academy of Sciences USA 46, 570-579.

Shannon, C.E. (1948). The mathematical theory of communication. Bell Systems Technical Journal 27, 379-423, 623-656. Reprinted in C.E. Shannon and W. Weaver (Eds.), The Mathematical Theory of Communication. Urbana and Chicago: Univeristy of Illinois Press (1963), pp. 30-125.

Timpson, C.G. (2000). Information and the Turing Principle: Some Philosophical Considerations. BPhil. Thesis, University of Oxford.\\ 
http://users.ox.ac.uk/~quee0776/thesis.html.

Timpson, C.G. (2002) The Applicability of the Shannon Information in Quantum Mechanics and Zeilinger's Foundational Principle. PSA 2002 Contributed Papers, forthcoming. PITT-PHIL-SCI-00000710.

Uffink, J.B.M (1990). Measures of Uncertainty and the Uncertainty Principle. Unpublished doctoral dissertation, University of Utrecht.

Wehrl, A. (1978). General properties of entropy. Reviews of Modern Physics 50(2), 221-260.

Wheeler, J.A. (1990). Information, Physics, Quantum: The Search for Links. In W. Zurek (Ed.), Complexity, Entropy and the Physics of Information. Redwood City, California: Addison-Wesley.

Wichmann, E.H. (1963). Density Matrices Arising from Incomplete Measurements. Journal of Mathematical Physics 4(7), 884-896.

Wootters, W.K. and Fields, B.D. (1989). Optimal State-Determination by Mutually Unbiased Measurements. Annals of Physics (New York) 191, 363.

Zeilinger, A. (1999). A Foundational Principle for Quantum Mechanics. Foundations of Physics 29(4), 631-643.

\end{document}